# Analysis of $\Lambda_b \to \Lambda \ell^+ \ell^-$ rare decays in a non-universal $Z'$ model

D. Banerjee[+] and S. Sahoo[*]

Department of Physics, National Institute of Technology,

Durgapur-713209, West Bengal, India

[+]E-mail: rumidebika@gmail.com

[*]E-mail: sukadevsahoo@yahoo.com

**Abstract:**

We investigate the rare baryonic $\Lambda_b \to \Lambda \ell^+ \ell^-$ decays in a non-universal $Z'$ model, which is one of the well-motivated extensions of the standard model (SM). Considering the effects of $Z'$-mediated flavour-changing neutral currents (FCNCs) we analyse the differential decay rate, forward-backward asymmetries and lepton polarisation asymmetries for the $\Lambda_b \to \Lambda \ell^+ \ell^-$ decays. We find significant deviations from their SM predictions, which could indicate new physics arising from the $Z'$ gauge boson.

## 1. Introduction

Rare baryonic decays $\Lambda_b \to \Lambda \ell^+ \ell^- \ (\ell = e, \mu, \tau)$ induced by flavour-changing neutral current (FCNC) occur at loop level in the standard model (SM) [1]. These decays can provide useful information about the parameters of the SM and also offer the possibility of searching for new physics (NP) beyond the SM. Joint efforts at hadron colliders and B factories have contributed much data of unprecedented precision in this sector [2-8]. The predictions based on the SM are in almost perfect agreement with the experimental findings of different particle colliders from all over the world. But in recent years several experimental results in this sector have shown deviations from the SM values, including observation of a 3.7σ deviation in the angular observable $P_5'$ [4] of the $B \to K^* \mu^+ \mu^-$ mode, the violation of lepton universality in $B \to K \ell^+ \ell^-$ decays at the level of 2.6 σ [5], a considerable discrepancy in the decay rates of the $B \to K^* \ell^+ \ell^-$ processes [6] observed by the LHCb experiment, the observed discrepancy in the branching fraction ratio of exclusive $B \to K^* \ell^+ \ell^-$ decays and inclusive decays into dimuons over dielectrons in the full range of $q^2$ [7], the observation of 3.3σ deviation in the decay rate of the $B_s \to \phi \mu^+ \mu^-$ [8] process and many more. Though these deviations are not statistically sufficient enough to prove the presence of NP effects, these data have intimated several anomalies in B decays induced by FCNC processes $b \to s \ell^+ \ell^-$. This prompts study of the implication of these observations in the context of various NP models as well as in a model independent way. Therefore, the rare $\Lambda_b$ decays involving $b \to s \ell^+ \ell^-$ transition provide a suitable way to search for NP effects. The NP arises in these decays in two different ways: either by introducing a new component to the Wilson coefficients or by modifying the structure of the effective Hamiltonian, both of which are absent in the SM. These decays have been studied in the literature both in the SM and in various beyond the SM (BSM) scenarios such as the two-Higgs-doublet model (2HDM), minimal supersymmetric standard model (MSSM) etc. [9-19]. Recent statistical analysis of $\Lambda_b \to \Lambda \mu^+ \mu^-$ decay using lattice QCD inputs has been done in Ref. [20].

The theoretical study of inclusive decays is easy but their experimental detection is quite difficult. For exclusive decays the situation is opposite i.e. their experimental detection is easy but theoretical analysis is very difficult. One type of exclusive decay which is described at inclusive level by the $b \to s \ell^+ \ell^-$ transition is the baryonic $\Lambda_b \to \Lambda \ell^+ \ell^-$ $(\ell = e, \mu, \tau)$ decays. These decays are studied in the SM [21], in the supersymmetric model with and without R-parity [22-25], in the two-Higgs-doublet model [26] and in a model-independent way. In comparison with B meson decays, $\Lambda_b$ baryon decays contain some particular observables, involving the spin of the b quark. So, the number of degrees of freedom involved in the bound state of baryon is more, hence the study of $\Lambda_b \to \Lambda \ell^+ \ell^-$ decays is less explored in comparison with B meson decays. In this paper, we study the $\Lambda_b \to \Lambda \ell^+ \ell^- (\ell = e, \mu, \tau)$ decay modes in the non-universal $Z'$ boson model and estimate the differential decay rate, forward-backward asymmetry and lepton polarisation asymmetries in these decay modes. Theoretically, non-universal $Z'$ bosons exist in many extensions of the SM, for example grand unified theories (GUTs) such as SU (5) or $E_6$ model [27, 28], superstring theories and theories with large extra dimensions. A non-universal $Z'$ [29-32] boson is one of the most important theoretically constructed model to understand physics beyond the SM [30-35]. As the $Z'$ boson has not yet been discovered experimentally, its mass is not known exactly, but there are stringent limits imposed by several theoretical models. The $Z'$ mass is constrained by direct searches from different accelerators [36-37], which give a model-dependent lower bound around 500 GeV. In a study of $B$ meson decays with $Z'$-mediated flavor-changing neutral currents [38], the $Z'$-boson was studied in the mass range of a few hundred GeV to 1 TeV. Sahoo *et al.* [39] estimated the $Z'$ boson mass from $B_q^0 - \overline{B_q^0}$ mixing, giving a result in the range 1352–1665 GeV. Oda *et al.* [40] have predicted an upper bound on $Z'$ boson mass of $M_{Z'} \leq 6$ TeV in a classically conformal $U(1)'$ extended standard model. The ATLAS collaboration [41] sets the lower mass limit for the sequential standard model (SSM) $Z'_{SSM}$ as 1.90 TeV and ranges from $1.82 - 2.17$ TeV are excluded for a $Z'_{SFM}$ strong flavor model. Recently, the CMS collaboration [42] has searched for leptophobic $Z'$ bosons decaying into four-lepton final states in proton-proton collisions at $\sqrt{s} = 8$ TeV and obtained the lower limit on the $Z'$ boson mass as 2.5 TeV. In this paper, we study the $Z'$-boson with mass in the TeV range.

Flavour mixing can be induced at tree level in the up-type and/or down-type quark sectors after diagonalizing their mass matrices. Mixing between ordinary and exotic left-handed quarks induces Z-mediated FCNCs. The right-handed quarks $d_R, s_R$ and $b_R$ have different $U(1)'$ quantum numbers than exotic $q_R$ and their mixing will induce $Z'$-mediated FCNCs [38, 43-46] among the ordinary down quark types. Tree level FCNC interactions can also be induced by an additional $Z'$ boson in the up-type quark sector [47]. In the $Z'$ model [48], the FCNC $b - s - Z'$ coupling is related to the flavour-diagonal couplings $qqZ'$ in a predictive way, which is then used to obtain upper limits on the leptonic $\ell\ell Z'$ couplings. With FCNCs, both Z and $Z'$ boson contributes at tree level and will interfere with the SM contributions [45-47, 49]. Hence, it is possible to study $\Lambda_b \to \Lambda \ell^+ \ell^-$ rare decay in the light of a non-universal $Z'$ model to explore beyond the SM. Since, at quark level the $\Lambda_b$ baryonic and $B$ mesonic decays are induced by the same mechanism, we can independently test our understanding of quark-hadron dynamics and investigate CP-asymmetry parameters with the help of $\Lambda_b \to \Lambda \ell^+ \ell^-$ rare decays by combining with experimental data from the mesonic sector.

This paper is organised as follows. In Section 2, we present the effective Hamiltonian responsible for the $b \to s \ell^+ \ell^-$ transitions and the matrix element for the decay modes $\Lambda_b \to \Lambda \ell^+ \ell^- (\ell = e, \mu, \tau)$ with their decay parameters in the SM. We also present the expressions of the forward backward asymmetry and lepton polarisation asymmetries for the same decay modes. In Section 3, we discuss the effect of the $Z'$ mediated FCNCs. We write the effective Hamiltonian for the $Z'$ part for $\Lambda_b \to \Lambda \ell^+ \ell^-$ decays following the modified Wilson coefficients $C_9$ and $C_{10}$. In Section 4, the numerical results of the physical observables - differential decay rate, forward-backward asymmetry and lepton polarisation asymmetries are discussed for the $\Lambda_b \to \Lambda \ell^+ \ell^- (\ell = e, \mu, \tau)$ decay modes in the non-universal $Z'$ boson model. Finally, we present our conclusions in Section 5.

## 2. $\Lambda_b \to \Lambda \ell^+ \ell^-$ decay in the standard model

At quark level the decay process $\Lambda_b \to \Lambda \ell^+ \ell^-$ is governed by the $b \to s \ell^+ \ell^-$ transition. In the SM the effective Hamiltonian responsible for the $b \to s \ell^+ \ell^-$ transition at the $O(m_b)$ scale is calculated by neglecting the doubly Cabbibo suppressed contributions [50, 51]. The matrix element of the $b \to s \ell^+ \ell^-$ process contains terms describing the virtual effects induced by $t\bar{t}, c\bar{c}$ and $u\bar{u}$ loops which are proportional to $V_{tb}V_{ts}^*, V_{cb}V_{cs}^*$ and $V_{ub}V_{us}^*$ respectively. Due to the unitarity of the CKM matrix and neglecting $V_{ub}V_{us}^*$ in comparison to $V_{tb}V_{ts}^*$ and $V_{cb}V_{cs}^*$, the matrix element of $b \to s \ell^+ \ell^-$ will contain only one independent CKM factor $V_{tb}V_{ts}^*$. The effective Hamiltonian describing the $\Lambda_b \to \Lambda \ell^+ \ell^-$ decay process is given as [50, 52]:

$$H_{eff} = \frac{G_F \alpha}{2\sqrt{2}\pi} V_{tb}V_{ts}^* [\ C_9^{eff}\left(\bar{s}\gamma_\mu(1-\gamma_5)b\right)\left(\bar{\ell}\gamma^\mu \ell\right) + C_{10}^{SM}\left(\bar{s}\gamma_\mu(1-\gamma_5)b\right)\left(\bar{\ell}\gamma^\mu\gamma_5 \ell\right)$$
$$- 2C_7^{eff} m_b \left(\bar{s} i \sigma_{\mu\nu} \frac{q^\mu}{q^2}(1+\gamma_5)b\right)\left(\bar{\ell}\gamma^\mu \ell\right)\ ],$$

(1)

where $G_F$ is the Fermi coupling constant, $\alpha$ is the electromagnetic coupling constant, $q$ is the momentum transferred to the lepton pair, which is the sum of the momenta of the $\ell^+$ and $\ell^-$ i.e. $q = p_+ + p_-$, and $C_7^{eff}$, $C_9^{SM}$ and $C_{10}^{SM}$ are Wilson coefficients evaluated at energy scale $\mu$ ($\mu = m_b$) and are given as [53],

$$C_7^{eff} = -0.308, \quad C_9^{SM} = 4.154, \quad C_{10}^{SM} = -4.261. \tag{2}$$

Considering long distance effects, a perturbative part coming from one loop matrix elements of the four quark operators ($Y(s)$) [26, 54] and a resonance part ($C_9^{res}$) [14, 55-57] due to the long distance resonance effect are introduced in the coefficient $C_9^{eff}$. Hence, it can be written as:

$$C_9^{eff} = C_9 + Y(s) + C_9^{res}. \tag{3}$$

The amplitude of the exclusive decay $\Lambda_b \to \Lambda \ell^+ \ell^-$ is obtained by sandwiching $H_{eff}$ for the $b \to s \ell^+ \ell^-$ transition between initial and final baryon states, i.e. $\langle \Lambda | H_{eff} | \Lambda_b \rangle$. The matrix elements of the various hadronic currents between the initial $\Lambda_b$ and the final $\Lambda$ baryon can be derived in terms of the form factors, as discussed in detail in [58]:

$$\langle \Lambda | \bar{s} \gamma_\mu b | \Lambda_b \rangle = \bar{u}_\Lambda \left[ f_1 \gamma_\mu + i f_2 \sigma_{\mu\nu} p^\nu + f_3 p_\mu \right] u_{\Lambda_b},$$

$$\langle \Lambda | \bar{s} \gamma_\mu \gamma_5 b | \Lambda_b \rangle = \bar{u}_\Lambda \left[ g_1 \gamma_\mu \gamma_5 + i g_2 \sigma_{\mu\nu} \gamma_5 p^\nu + g_3 \gamma_5 p_\mu \right] u_{\Lambda_b},$$

$$\langle \Lambda | \bar{s} i \sigma_{\mu\nu} p^\nu b | \Lambda_b \rangle = \bar{u}_\Lambda \left[ f_1^T \gamma_\mu + i f_2^T \sigma_{\mu\nu} p^\nu + f_3^T p_\mu \right] u_{\Lambda_b},$$

$$\langle \Lambda | \bar{s} i \sigma_{\mu\nu} \gamma_5 p^\nu b | \Lambda_b \rangle = \bar{u}_\Lambda \left[ g_1^T \gamma_\mu \gamma_5 + i g_2^T \sigma_{\mu\nu} \gamma_5 p^\nu + g_3^T \gamma_5 p_\mu \right] u_{\Lambda_b}, \tag{4}$$

where $f_i$ and $g_i$ are the various form factors which are functions of $q^2$. The decay process $\Lambda_b \to \Lambda$ is studied based on the heavy quark effective theory (HQET) in [59]. Using the heavy quark symmetry limit, the number of independent form factors is reduced to two and the matrix elements of all hadronic currents, irrespective of their Dirac structure, can be written as

$$\langle \Lambda(p_\Lambda) | \bar{s} \Gamma b | \Lambda_b(p_{\Lambda_b}) \rangle = \bar{u}_\Lambda \left[ F_1(p^2) + \slashed{v} F_2(p^2) \right] \Gamma u_{\Lambda_b}, \tag{5}$$

where $\Gamma$ is the product of the Dirac matrices and $F_{1,2}$ are the form factors. The relations among these two sets of form factors are given as [22-25, 60]:

$$g_1 = f_1 = f_2^T = g_2^T = F_1 + \sqrt{r}\, F_2, \qquad g_2 = f_2 = g_3 = f_3 = \frac{F_2}{m_{\Lambda_b}},$$

$$g_3^T = \frac{F_2}{m_{\Lambda_b}}(m_{\Lambda_b} + m_\Lambda), \quad f_3^T = -\frac{F_2}{m_{\Lambda_b}}(m_{\Lambda_b} - m_\Lambda), \quad f_1^T = g_1^T = \frac{F_2}{m_{\Lambda_b}} q^2. \tag{6}$$

where $r = m_\Lambda^2 / m_{\Lambda_b}^2$. The form factors $F_1$ and $F_2$ for the $\Lambda_b \to \Lambda \ell^+ \ell^-$ decay are calculated in the QCD sum rule approach combined with heavy quark symmetry in [21-25, 60] and the transition amplitude can be written as [58]:

$$M(\Lambda_b \to \Lambda \ell^+ \ell^-) = \frac{G_F \alpha}{\sqrt{2}\pi} V_{tb} V_{ts}^* \times \begin{bmatrix} \bar{\ell}\gamma_\mu \ell \left\{ \bar{u}_\Lambda \left[ \gamma^\mu (A_1 P_R + B_1 P_L) + i\sigma^{\mu\nu} q_\nu (A_2 P_R + B_2 P_L) \right] u_{\Lambda_b} \right\} \\ + \bar{\ell}\gamma_\mu \gamma_5 \ell \left\{ \bar{u}_\Lambda \begin{bmatrix} \gamma^\mu (D_1 P_R + E_1 P_L) + i\sigma^{\mu\nu} q_\nu (D_2 P_R + E_2 P_L) \\ + q^\mu (D_3 P_R + E_3 P_L) \end{bmatrix} u_{\Lambda_b} \right\} \end{bmatrix}$$
(7)

where the parameters $A_i$, $B_i$ and $D_j$, $E_j$ ($i = 1, 2$ and $j = 1, 2, 3$) are defined as

$$A_i = \frac{1}{2} C_9^{eff} (f_i - g_i) - \frac{2 C_7^{eff} m_b}{q^2} (f_i^T + g_i^T),$$

$$B_i = \frac{1}{2} C_9^{eff} (f_i + g_i) - \frac{2 C_7^{eff} m_b}{q^2} (f_i^T - g_i^T),$$

$$D_j = \frac{1}{2} C_{10} (f_j - g_j), \quad E_j = \frac{1}{2} C_{10} (f_j + g_j). \tag{8}$$

The double partial decay rates for $\Lambda_b \to \Lambda \ell^+ \ell^-$ ($\ell = e, \mu, \tau$) can be obtained from the transition amplitude [Equation (7)] as:

$$\frac{d^2 \Gamma}{d\hat{s}\, dz} = \frac{G_F^2 \alpha^2 |V_{tb} V_{ts}^*|^2}{2^{12} \pi^5} m_{\Lambda_b} v_\ell \sqrt{1 - \frac{4 m_\ell^2}{q^2}} \sqrt{\lambda(1, r, \hat{s})}\, K(\hat{s}, z), \tag{9}$$

where, $\hat{s} = s/m_{\Lambda_b}^2$, $s = q^2$, $z = \cos\theta$, the angle between $p_{\Lambda_b}$ and $p_+$ in the center of mass frame of $\ell^+ \ell^-$ pair and $\lambda(a, b, c) = a^2 + b^2 + c^2 - 2(ab + bc + ca)$ is the usual triangle function. The function $K(\hat{s}, z)$ is given as

$$K(\hat{s}, z) = K_0(\hat{s}) + z K_1(\hat{s}) + z^2 K_2(\hat{s}), \tag{10}$$

where, $K_0(\hat{s}) = 32 m_l^2 m_{\Lambda_b}^4 \hat{s}(1+r-\hat{s})(|D_3|^2 + |E_3|^2) + 64 m_l^2 m_{\Lambda_b}^3 (1-r-\hat{s}) \text{Re}(D_1^* E_3 + D_3 E_1^*)$

$+ 64 m_{\Lambda_b}^2 \sqrt{r} (6m_l^2 - \hat{s} m_{\Lambda_b}^2) \text{Re}(D_1^* E_1)$

$+ 64 m_l^2 m_{\Lambda_b}^3 \sqrt{r} \times \left[ 2 m_{\Lambda_b} \hat{s} \, \text{Re}(D_3^* E_3) + (1-r+\hat{s}) \text{Re}(D_1^* D_3 + E_1^* E_3) \right]$

$+ 32 m_{\Lambda_b}^2 (2m_l^2 + \hat{s} m_{\Lambda_b}^2) \times \begin{bmatrix} (1-r+\hat{s}) m_{\Lambda_b} \sqrt{r} \, \text{Re}(A_1^* A_2 + B_1^* B_2) - m_{\Lambda_b}(1-r-\hat{s}) \text{Re}(A_1^* B_2 + A_2^* B_1) \\ -2\sqrt{r} \{ \text{Re}(A_1^* B_1) + m_{\Lambda_b}^2 \hat{s} \, \text{Re}(A_2^* B_2) \} \end{bmatrix}$

$+ 8 m_{\Lambda_b}^2 \left[ 4 m_l^2 (1+r-\hat{s}) + m_{\Lambda_b}^2 \{(1-r)^2 - \hat{s}^2\} \right] \times \left( |A_1|^2 + |B_1|^2 \right)$

$+ 8 m_{\Lambda_b}^4 \left[ 4 m_l^2 \{ \lambda + (1+r-\hat{s})\hat{s} \} + m_{\Lambda_b}^2 \hat{s} \{(1-r)^2 - \hat{s}^2\} \right] \times \left( |A_2|^2 + |B_2|^2 \right)$

$- 8 m_{\Lambda_b}^2 \left[ 4 m_l^2 (1+r-\hat{s}) - m_{\Lambda_b}^2 \{(1-r)^2 - \hat{s}^2\} \right] \times \left( |D_1|^2 + |E_1|^2 \right)$

$+ 8 m_{\Lambda_b}^5 \hat{s} v_l^2 \times \begin{bmatrix} -8 m_{\Lambda_b} \hat{s} \sqrt{r} \, \text{Re}(D_2^* E_2) + 4(1-r+\hat{s})\sqrt{r} \, \text{Re}(D_1^* D_2 + E_1^* E_2) \\ -4(1-r-\hat{s}) \text{Re}(D_1^* E_2 + D_2^* E_1) + m_{\Lambda_b} \{(1-r)^2 - \hat{s}^2\} \{ |D_2|^2 + |E_2|^2 \} \end{bmatrix}$,

(11)

$K_1(\hat{s}) = -16 m_{\Lambda_b}^4 \hat{s} v_l \sqrt{\lambda} \{ 2 \text{Re}(A_1^* D_1) - 2 \text{Re}(B_1^* E_1) + 2 m_{\Lambda_b} \text{Re}(B_1^* D_2 - B_2^* D_1 + A_2^* E_1 - A_1^* E_2) \}$

$+ 32 m_{\Lambda_b}^5 \hat{s} v_l \sqrt{\lambda} \{ m_{\Lambda_b}(1-r) \text{Re}(A_2^* D_2 - B_2^* E_2) + \sqrt{r} \, \text{Re}(A_2^* D_1 + A_1^* D_2 - B_2^* E_1 - B_1^* E_2) \}$

(12)

and $\quad K_2(\hat{s}) = 8 m_{\Lambda_b}^6 \hat{s} v_l^2 \lambda \left[ |A_2|^2 + |B_2|^2 + |D_2|^2 + |E_2|^2 \right]$

$\quad - 8 m_{\Lambda_b}^4 v_l^2 \lambda \left[ |A_1|^2 + |B_1|^2 + |D_1|^2 + |E_1|^2 \right]$, (13)

where $\lambda$ is the short form for $\lambda(1, r, \hat{s})$. Now integrating Equation (9) w. r. t. the angular dependent parameter z, we can get,

$$\left( \frac{d\Gamma}{d\hat{s}} \right)_0 = \frac{G_F^2 \alpha^2 |V_{tb} V_{ts}^*|^2}{2^{11} \pi^5 m_{\Lambda_b}} \sqrt{1 - \frac{4 m_l^2}{q^2}} \sqrt{\lambda(1, r, \hat{s})} \left[ K_0(\hat{s}) + \frac{1}{3} K_2(\hat{s}) \right],$$
(14)

The forward-backward asymmetry is defined as,

$$A_{FB}(\hat{s}) = \frac{\int_0^1 \frac{d\Gamma}{d\hat{s} dz} dz - \int_{-1}^0 \frac{d\Gamma}{d\hat{s} dz} dz}{\int_0^1 \frac{d\Gamma}{d\hat{s} dz} dz + \int_{-1}^0 \frac{d\Gamma}{d\hat{s} dz} dz}$$
(15)

Hence, we can derive $A_{FB}$ using Equation (11-15) in the following form,

$$A_{FB}(\hat{s}) = \frac{K_1(\hat{s})}{K_0(\hat{s}) + K_2(\hat{s})/3} \ . \qquad (16)$$

When the final $\Lambda$ baryon is polarised, the lepton polarization components $P_i (i = L, N, T)$ are defined as:

$$P_i(\hat{s}) = \frac{\frac{d\Gamma}{d\hat{s}}(\hat{\eta} = \hat{e}_i) - \frac{d\Gamma}{d\hat{s}}(\hat{\eta} = -\hat{e}_i)}{\frac{d\Gamma}{d\hat{s}}(\hat{\eta} = \hat{e}_i) + \frac{d\Gamma}{d\hat{s}}(\hat{\eta} = -\hat{e}_i)} \ , \qquad (17)$$

where the $\hat{e}_i$'s are the orthogonal unit vectors along longitudinal, normal and transverse components of the $\ell^+$ polarization and $\hat{\eta}$ is the unit vector used to represent the four spin vector, along the spin in its rest frame as:

$$s_+^0 = \frac{\vec{p}_+ \cdot \hat{\eta}}{m_\ell}, \quad \vec{s}_+ = \hat{\eta} + \frac{s_+^0}{E_{\ell^+} + m_\ell} \vec{p}_+ \qquad (18)$$

The observables $P_L, P_N, P_T$ correspond to longitudinal, normal and transverse polarization asymmetries respectively. Among these observables, $P_L$ and $P_T$ are $P$-odd, $T$-even whereas $P_N$ is $P$-even, $T$-odd. The observables $P_L, P_N, P_T$ are given as [58]:

$$P_L(\hat{s}) = \frac{16 m_{\Lambda_b}^2 \sqrt{\lambda}}{K_0(\hat{s}) + K_2(\hat{s})/3}$$

$$\times [\ 8 m_\ell^2 m_{\Lambda_b} \times \left( \mathrm{Re}(D_1^* E_3 - D_3^* E_1) + \sqrt{r}\, \mathrm{Re}(D_1^* D_3 - E_1^* E_3) \right)$$

$$- 4 m_\ell^2 m_{\Lambda_b}^2 \hat{s} \left( |D_3|^2 - |E_3|^2 \right) - 4 m_{\Lambda_b} \left( 2 m_\ell^2 + m_{\Lambda_b}^2 \hat{s} \right) \mathrm{Re}(A_1^* B_2 - A_1^* B_2)$$

$$- \frac{4}{3} m_{\Lambda_b}^3 \hat{s} v_\ell^2 \left( 3 \mathrm{Re}(D_1^* E_2 - D_2^* E_1) + \sqrt{r}\, \mathrm{Re}(D_1^* D_2 - E_1^* E_2) \right)$$

$$- \frac{4}{3} m_{\Lambda_b} \sqrt{r} \left( 6 m_\ell^2 + m_{\Lambda_b}^2 \hat{s} v_\ell^2 \right) \mathrm{Re}(A_1^* A_2 - B_1^* B_2)$$

$$- \frac{2}{3} m_{\Lambda_b}^4 \hat{s}(2 - 2r + \hat{s}) v_\ell^2 \left( |D_2|^2 - |E_2|^2 \right) + \left( 4 m_\ell^2 + m_{\Lambda_b}^2 (1 - r + \hat{s}) \right) \left( |A_1|^2 - |B_1|^2 \right)$$

$$- \left( 4 m_\ell^2 - m_{\Lambda_b}^2 (1 - r + \hat{s}) \right) \left( |D_1|^2 - |E_1|^2 \right)$$

$$- \frac{1}{3} m_{\Lambda_b}^2 (1 - r - \hat{s}) v_\ell^2 \left( |A_1|^2 - |B_1|^2 + |D_1|^2 - |E_1|^2 \right)$$

$$- \frac{1}{3} m_{\Lambda_b}^2 \left[ 12 m_\ell^2 (1 - r) + m_{\Lambda_b}^2 \hat{s} (3(1 - r + \hat{s}) + v_\ell^2 (1 - r - \hat{s})) \right] \times \left( |A_2|^2 - |B_2|^2 \right) ] \qquad (19)$$

$$P_N(\hat{s}) = \frac{8\pi\, m_{\Lambda_b}^3 v_\ell \sqrt{\lambda}}{K_0(\hat{s}) + K_2(\hat{s})/3} \times \Big[ -2m_{\Lambda_b}^2 (1-r+\hat{s}) \times \sqrt{r}\, \text{Re}(A_1^* D_1 - B_1^* E_1)$$

$$+ 4m_{\Lambda_b}^2 \hat{s}\sqrt{r}\, \text{Re}(A_1^* E_2 + A_2^* E_1 + B_1^* D_2 + B_2^* D_1) - 2m_{\Lambda_b}^3 \hat{s}\sqrt{r}(1-r+\hat{s})\text{Re}(A_2^* D_2 + B_2^* D_2)$$

$$+ 2m_{\Lambda_b}(1-r-\hat{s}) \times \big(\text{Re}(A_1^* E_1 + B_1^* D_1) + m_{\Lambda_b}^2 \hat{s}\, \text{Re}(A_2^* E_2 + B_2^* D_2)\big)$$

$$- m_{\Lambda_b}^2 ((1-r)^2 - \hat{s}^2) \times \text{Re}(A_1^* D_2 + A_2^* D_1 + B_1^* E_2 + B_2^* E_1) \Big]. \qquad (20)$$

$$P_T(\hat{s}) = \frac{8\pi\, m_{\Lambda_b}^3 v_\ell \sqrt{\hat{s}\lambda}}{K_0(\hat{s}) + K_2(\hat{s})/3} \times \Big[\, m_{\Lambda_b}^2 (1-r-\hat{s})\, \big(\text{Im}(A_2^* D_1 - A_1^* D_2) - \text{Im}(B_2^* E_1 - B_1^* E_2)\big)$$

$$+ 2m_{\Lambda_b}\, \text{Im}(A_1^* E_1 - B_1^* D_1) - m_{\Lambda_b}^2 \hat{s}\, \text{Im}(A_2^* E_2 - B_2^* D_2)\,\Big]. \qquad (21)$$

### 3. Contribution of $Z'$- mediated FCNC in $\Lambda_b \to \Lambda \ell^+ \ell^-$ decays

A non-universal $Z'$ model could be naturally derived from grand unified theories (GUT's), superstring theories, $E_6$ models and so on, but experimentally the $Z'$ boson has not yet been discovered. One convenient process to get the $Z'$ boson is to include an additional $U(1)'$ gauge symmetry in the SM. In the non-universal $Z'$ model, the FCNC transitions for the $b \to s \ell^+ \ell^-$ process are induced at tree level due to the presence of a non-diagonal chiral coupling matrix. The detailed formulation of this model is discussed in [32]. By neglecting the $Z - Z'$ mixing and considering that the couplings of righ- handed quark flavors with the $Z'$ boson are diagonal, the $Z'$ part of the effective Hamiltonian for $\Lambda_b \to \Lambda \ell^+ \ell^-$ decays is written as [47-48, 61-67]:

$$H_{eff}^{Z'} = -\frac{2G_F}{\sqrt{2}\pi} V_{tb} V_{ts}^* \left[ \frac{B_{sb}^L S_{\ell\ell}^L}{V_{tb} V_{ts}^*} \bar{s}\gamma_\mu(1-\gamma_5) b\, \bar{\ell}\gamma^\mu(1-\gamma_5)\ell + \frac{B_{sb}^L S_{\ell\ell}^R}{V_{tb} V_{ts}^*} \bar{s}\gamma_\mu(1-\gamma_5) b\, \bar{\ell}\gamma^\mu(1+\gamma_5)\ell \right] \qquad (22)$$

In the above expression, $B_{sb}^L = |B_{sb}| e^{-i\phi_{sb}}$, which corresponds to the off-diagonal left-handed couplings of quarks with the $Z'$ boson and $\phi_{sb}$ is the new weak phase. In a compact manner, the above equation can be rewritten as;

$$H_{eff}^{Z'} = -\frac{4G_F}{\sqrt{2}} V_{tb} V_{ts}^* \left[ \Lambda_{sb} C_9^{Z'} O_9 + \Lambda_{sb} C_{10}^{Z'} O_{10} \right], \qquad (23)$$

where $\Lambda_{sb} = \dfrac{4\pi e^{-i\phi_{sb}}}{\alpha V_{tb}V_{ts}^*}$, (24)

$$C_9^{Z'} = |B_{sb}|S_{LL}, \quad C_{10}^{Z'} = |B_{sb}|D_{LL},$$ (25)

and $S_{LL} = S_{\ell\ell}^L + S_{\ell\ell}^R, \quad D_{LL} = S_{\ell\ell}^R - S_{\ell\ell}^L.$ (26)

Here, $S_{\ell\ell}^L$ and $S_{\ell\ell}^R$ represent the couplings of the $Z'$ boson with the left- and right-handed leptons respectively. The contributions to the decay $\Lambda_b \to \Lambda \ell^+ \ell^-$ mainly come from the Wilson coefficients $C_9$ and $C_{10}$, and corresponding operators. In this $Z'$ model the operator basis remain the same as in the SM and the contribution of $Z'$ only modifies the Wilson coefficients $C_9$ and $C_{10}$. Hence, to include the $Z'$ contributions, it is sufficient to make the following replacements in the formalism relevant to $\Lambda_b \to \Lambda \ell^+ \ell^-$ decay:

$$C_9^{SM+Z'} = C_9^{SM} + \dfrac{4\pi e^{-i\phi_{sb}}}{\alpha V_{tb}V_{ts}^*}|B_{sb}|S_{LL}$$ (27)

$$C_{10}^{SM+Z'} = C_{10}^{SM} + \dfrac{4\pi e^{-i\phi_{sb}}}{\alpha V_{tb}V_{ts}^*}|B_{sb}|D_{LL}$$ (28)

In this model, the new physics contributions to forward-backward asymmetry and other polarisation asymmetries for $\Lambda_b \to \Lambda \ell^+ \ell^-$ decay are analysed in the light of the above modifications in Section 4.

## 4. Numerical Analysis

In this section, we discuss the forward-backward asymmetries and lepton polarisation asymmetries for $\Lambda_b \to \Lambda \ell^+ \ell^- \;(\ell = e, \mu, \tau)$ decays in the frame work of the non-universal $Z'$ model and investigate the scenario of NP. The numerical values of the input parameters [7] used are collectively given in Table - 1. The dependence of the form factors for $f_{1,2,3}$, $g_{1,2,3}$, $f_{2,3}^T$ and $g_{2,3}^T$ on $q^2$ in the light–cone QCD sum rules approach can be parameterized as [68]:

$$f_i(q^2)[g_i(q^2)] = \dfrac{a}{1 - q^2/m_{fit}^2} + \dfrac{b}{(1 - q^2/m_{fit}^2)^2},$$ (29)

whereas the form factors $f^T_1$ and $g^T_1$ are of the form

$$f_1^T(q^2)[g_1^T(q^2)] = \dfrac{a}{1 - q^2/m_{fit}'^2} + \dfrac{b}{(1 - q^2/m_{fit}''^2)^2}.$$ (30)

The parameters appearing in the fit function of the form factors are summarised in Table - 2. To evaluate different observables in the $Z'$ model we need to fix the numerical values of the parameters $|B_{sb}|$, $\varphi_{sb}$, $S_{LL}$ and $D_{LL}$ for the $Z'$ couplings. These values, however, are strictly constrained from B meson mixing and different inclusive as well as exclusive decays of B mesons. The values are collected from Refs. [69-85] and are encapsulated in Table - 3 for three different scenarios $S_1$, $S_2$ and $S_3$, where $S_1$ and $S_2$ correspond to two different fitting values for $B_s - \bar{B}_s$ mixing data from the UTfit Collaboration [70-76] and $S_3$ is obtained from the analysis of three different *B* meson decays [79-82].

Table-1: Numerical values of the input parameters [7].

| Parameters | Value |
|---|---|
| $M_{\Lambda_b}$ | 5.620 GeV |
| $M_{\Lambda}$ | 1.115 GeV |
| $m_b$ | 4.28 GeV |
| $m_e$ | 0.510 x $10^{-3}$ GeV |
| $m_\mu$ | 0.105 GeV |
| $m_\tau$ | 1.77 GeV |
| $G_F$ | 1.17 x $10^{-5}$ GeV$^{-2}$ |
| $\tau_{\Lambda_b}$ | 1.383 x $10^{-12}$ s |
| $\alpha$ | 1/137 |
| $|V_{tb}V_{ts}^*|$ | 45 x $10^{-3}$ |

Table-2: Parameters appearing in the fit function of the form factors [68].

| Fit parameter | a | b | $m^2_{fit}$ |
|---|---|---|---|
| $f_1$ | -0.046 | 0.368 | 39.10 |
| $f_2$ | 0.0046 | -0.017 | 26.37 |
| $f_3$ | 0.006 | -0.021 | 22.99 |
| $g_1$ | -0.220 | 0.538 | 48.70 |
| $g_2$ | 0.005 | -0.018 | 26.93 |
| $g_3$ | 0.035 | -0.050 | 24.26 |
| $f_2^T$ | -0.131 | 0.426 | 45.70 |
| $f_3^T$ | -0.046 | 0.102 | 28.31 |
| $g_2^T$ | -0.369 | 0.664 | 59.37 |
| $g_3^T$ | -0.026 | -0.075 | 23.73 |
| Fit parameter | C | $m'^2_{fit}$ | $m''^2_{fit}$ |
| $f_1^T$ | -1.191 | 23.81 | 59.96 |
| $g_1^T$ | -0.653 | 24.15 | 48.52 |

Table-3: Numerical values of the $Z'$ coupling parameters [69-85].

| Scenario | $\|B_{sb} \times 10^{-3}\|$ | $\phi_{sb}$ in degree | $S_{LL} \times 10^{-2}$ | $D_{LL} \times 10^{-2}$ |
|---|---|---|---|---|
| S1 | $1.09 \pm 0.22$ | $72 \pm 7$ | $-2.8 \pm 3.9$ | $-6.7 \pm 2.6$ |
| S2 | $2.20 \pm 0.15$ | $82 \pm 4$ | $-1.2 \pm 1.4$ | $-2.5 \pm 0.9$ |
| S3 | $4.0 \pm 1.5$ | $150 \pm 10$ or $-150 \pm 10$ | 0.8 | -2.6 |

In our calculation, we have considered the maximum values of the $Z'$ parameters from the three given scenarios in order to observe the maximum influence of the $Z'$ boson on the different asymmetry observables. So we construct three sets from the three scenarios of the numerical values of coupling parameters, as follows:

Set-I:

Within the range of coupling parameters for scenarios $S_1$ listed in Table-3, we have grouped the maximum values in this set to get the magnified impact of $Z'$ boson, i.e.

$|B_{sb}| = 1.31 \times 10^{-3}$, $S_{LL} = 1.1 \times 10^{-2}$, $D_{LL} = -4.1 \times 10^{-2}$, $\phi_{sb} = (-72 \pm 7)°$.

Set-II:

Within the range of coupling parameters for scenarios $S_2$ listed in Table-3, we have grouped the maximum values in this set to get the magnified impact of $Z'$ boson, i.e.

$|B_{sb}| = 2.53 \times 10^{-3}$, $S_{LL} = 0.2 \times 10^{-2}$, $D_{LL} = -1.6 \times 10^{-2}$, $\phi_{sb} = (-82 \pm 4)°$.

Set-III:

Within the range of coupling parameters for scenarios $S_3$ listed in Table-3, we have grouped the maximum values in this set to get the magnified impact of $Z'$ boson, i.e.

$|B_{sb}| = 5.7 \times 10^{-3}$, $S_{LL} = 0.8 \times 10^{-2}$, $D_{LL} = -2.6 \times 10^{-2}$, $\phi_{sb} = (-150 \pm 10)°$.

Let us proceed further with all the numerical data discussed above. In the $Z'$ model, the new physics contribution to the asymmetry parameters are encoded in the modified Wilson coefficient. Therefore, we investigate the variation of asymmetry observables for $\Lambda_b \to \Lambda \ell^+ \ell^-$ $(\ell = e, \mu, \tau)$ decays with different values of $Z'$ coupling parameter within the kinematically accessible physical range of $\hat{s}$.

For $\Lambda_b \to \Lambda e^+ e^-$ and $\Lambda_b \to \Lambda \mu^+ \mu^-$ decays, $A_{FB}(\hat{s})$ initially decreases with increase in $\hat{s}$ and takes small negative value for $\hat{s}$ up to 0.1, then $A_{FB}(\hat{s})$ gradually increases almost equally in the three sets of coupling parameters in Fig. 1(a) and Fig. 1(b) respectively. In Fig. 1(c), $A_{FB}(\hat{s})$ prominently decreases and then increases with respect to $\hat{s}$ for $\Lambda_b \to \Lambda \tau^+ \tau^-$ decay. In Fig. 1 (c), the maximum variation is observed with Set-III, hence we can say that with the higher contribution of coupling parameters $A_{FB}(\hat{s})$ increases.

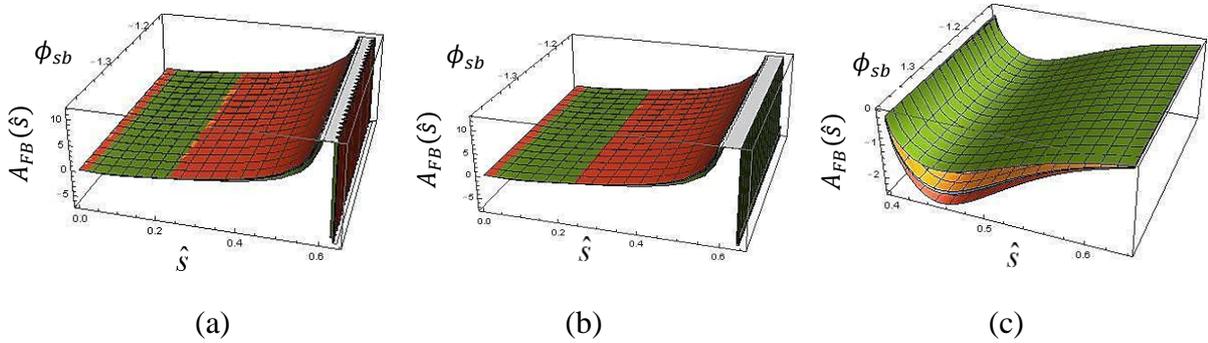

(a)          (b)          (c)

Fig. 1: Dependence of $A_{FB}(\hat{s})$ on $\hat{s}$ and $\phi_{sb}$ for the SM (red), Set-I (yellow), Set-II (blue), and Set-III (green) for the decays (a) $\Lambda_b \to \Lambda e^+ e^-$, (b) $\Lambda_b \to \Lambda \mu^+ \mu^-$ and (c) $\Lambda_b \to \Lambda \tau^+ \tau^-$.

We plot the differential decay rate (Eq$^n$ 14) for $\Lambda_b \to \Lambda \ell^+ \ell^-$ $(\ell = e, \mu, \tau)$ decays against $\hat{s}$ and $\phi_{sb}$ for Set-I (yellow), Set-II (blue) and Set-III (green), as depicted in Figs. 2(a), 2(b) and 2(c). It is observed that Set-III data contributes more to the differential decay rate for $\Lambda_b \to \Lambda \ell^+ \ell^-$ decays. Similarly, Figs. 3, 4 and 5 present the dependence of lepton polarization asymmetries $P_L$, $P_N$ and $P_T$ on $\hat{s}$ for $\Lambda_b \to \Lambda \ell^+ \ell^-$ decays respectively. For $\Lambda_b \to \Lambda e^+e^-$ and $\Lambda_b \to \Lambda\mu^+\mu^-$ decays, $P_L$ decreases in $Z'$ scenarios compare to the SM [ Fig. 3(a) and 3(b) ] whereas for the $\Lambda_b \to \Lambda\tau^+\tau^-$ decay, $P_L$ is enhanced in the three sets of $Z'$ coupling parameters compared to the SM values [ Fig. 3(c) ]. In Figs. 4(a), 4(b) and 4(c), the contribution of the $Z'$ boson to $P_N$ dominates over the SM value for $\Lambda_b \to \Lambda\ell^+\ell^-$ decays. Figures 5(a), 5(b) and 5(c) represent the variation of $P_T$ for $\Lambda_b \to \Lambda\ell^+\ell^-$ $(\ell = e, \mu, \tau)$ decays. For $\Lambda_b \to \Lambda e^+e^-$ and $\Lambda_b \to \Lambda\mu^+\mu^-$ decays, the $P_T$ decreases in $Z'$ scenarios compared to the SM [ Fig. 5(a) and 5(b) ], whereas for the $\Lambda_b \to \Lambda\tau^+\tau^-$ decay, $P_T$ is enhanced in the three sets of $Z'$ coupling parameters compared to the SM values [Fig. 5(c)]. The slope of the planes are similar for $\Lambda_b \to \Lambda e^+e^-$ and $\Lambda_b \to \Lambda\mu^+\mu^-$ decays, whereas the slope for the $\Lambda_b \to \Lambda\tau^+\tau^-$ decay is different. This may indicate the lepton non-universality, although it is absent in the SM.

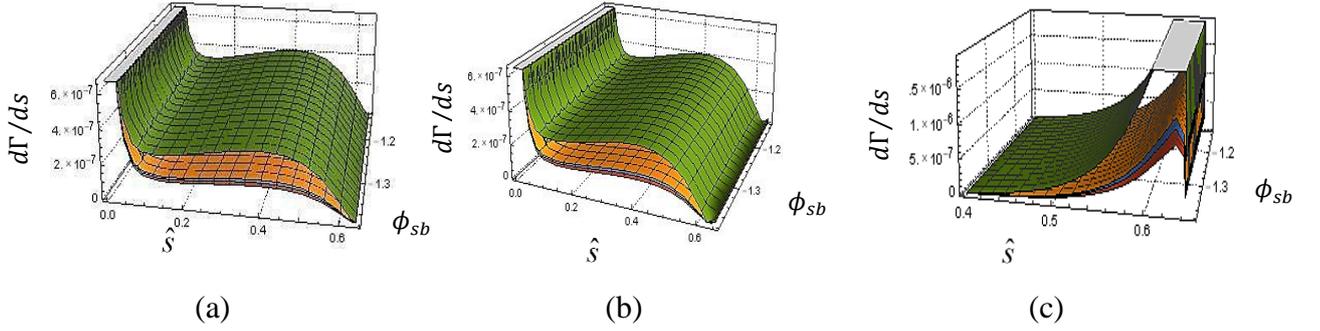

Fig.2: Dependence of differential decay rate on $\hat{s}$ and $\phi_{sb}$ for the SM (red), Set-I (yellow), Set-II (blue), and Set-III (green) for the decays (a) $\Lambda_b \to \Lambda e^+e^-$, (b) $\Lambda_b \to \Lambda\mu^+\mu^-$ and (c) $\Lambda_b \to \Lambda\tau^+\tau^-$.

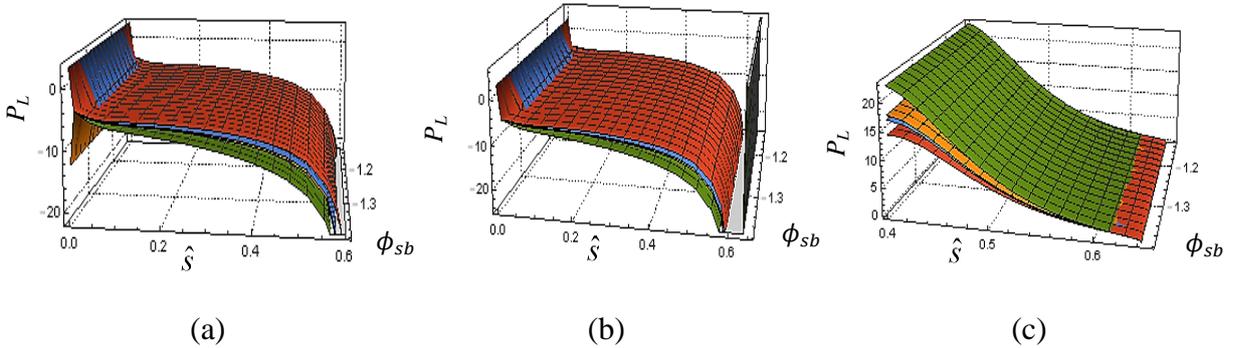

Fig.3: Dependence of longitudinal lepton polarization asymmetries ($P_L$) on $\hat{s}$ and $\phi_{sb}$ for the SM (red), Set-I (yellow), Set-II (blue) and Set-III (green) for the decays (a) $\Lambda_b \to \Lambda e^+e^-$, (b) $\Lambda_b \to \Lambda\mu^+\mu^-$ and (c) $\Lambda_b \to \Lambda\tau^+\tau^-$.

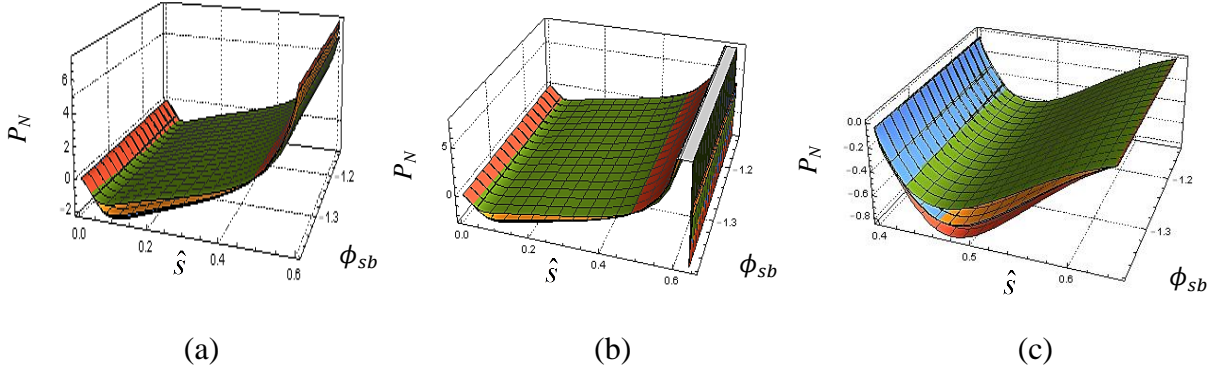

Fig.4: Dependence of normal lepton polarization asymmetries ($P_N$) on $\hat{s}$ and $\phi_{sb}$ for the SM (red), Set-I (yellow), Set-II (blue) and Set-III (green) for the decays (a) $\Lambda_b \to \Lambda e^+e^-$, (b) $\Lambda_b \to \Lambda\mu^+\mu^-$ and (c) $\Lambda_b \to \Lambda\tau^+\tau^-$.

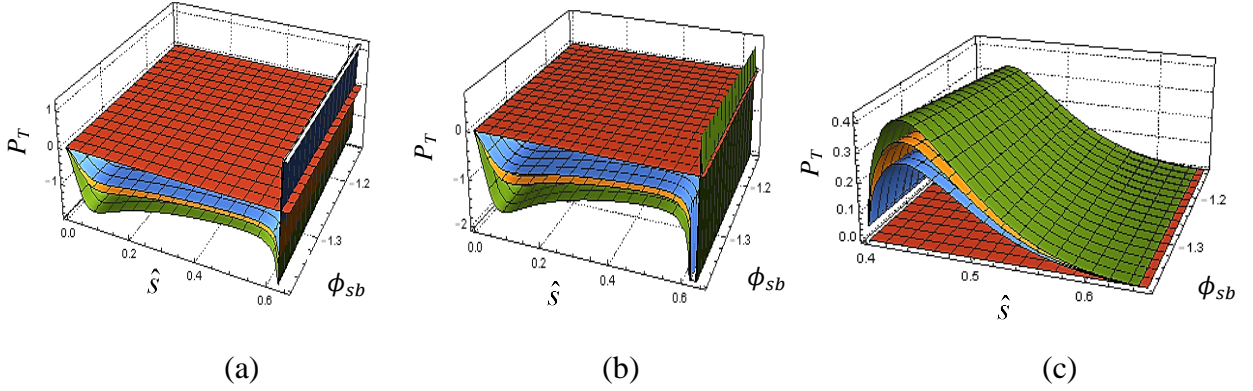

Fig.5: Dependence of transverse lepton polarization asymmetries ($P_T$) on $\hat{s}$ and $\phi_{sb}$ for the SM (red), Set-I (yellow), Set-II (blue) and Set-III (green) for the decays (a) $\Lambda_b \to \Lambda e^+e^-$, (b) $\Lambda_b \to \Lambda\mu^+\mu^-$ and (c) $\Lambda_b \to \Lambda\tau^+\tau^-$.

For $\hat{s} = 0.6$, $A_{FB}(\hat{s})$ is enhances significantly from that of the SM values with increasing values of $S_{LL}$ and $D_{LL}$ in $\Lambda_b \to \Lambda\mu^+\mu^-$ decay with $|B_{sb}| = 1.31 \times 10^{-3}$, $\phi_{sb} = -65°$ [Fig. 6 (a)]. Figure 6 (b) depicts a different picture for $\Lambda_b \to \Lambda\mu^+\mu^-$ decay in the high $\hat{s}$ region for $|B_{sb}| = 5.7 \times 10^{-3}$ and $\phi_{sb} = -140°$. It is observed that in the $Z'$ model, initially the values of $A_{FB}(\hat{s})$ are less than the SM prediction, but it gradually increases with increase in $Z'$ coupling parameters and finally cross over the SM value. Similar plots are presented in both Fig. 7(a) and 7(b) for $\Lambda_b \to \Lambda\tau^+\tau^-$ decay, showing that $A_{FB}(\hat{s})$ is significantly enhanced compared to the SM values with increasing $S_{LL}$ and $D_{LL}$. These plots present a clear distinction among the $Z'$ boson contribution to the forward-backward asymmetry and the SM values, which give signals for the existence of NP.

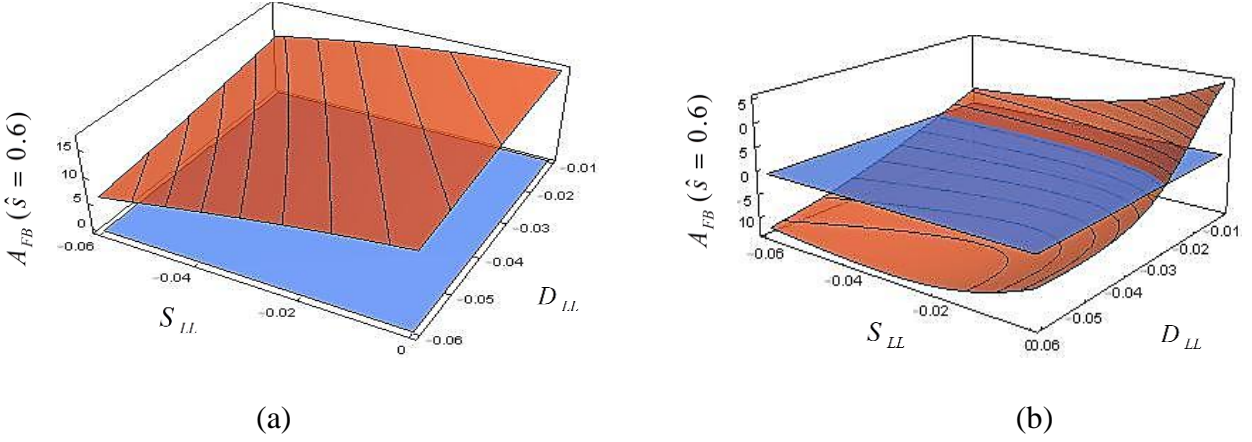

(a)          (b)

Fig.6: Dependence of $A_{FB}(\hat{s})$ on $S_{LL}$ and $D_{LL}$ at $\hat{s} = 0.6$ with (a) $|B_{sb}| = 1.31 \times 10^{-3}$, $\phi_{sb} = -65°$ and (b) $|B_{sb}| = 5.7 \times 10^{-3}$, $\phi_{sb} = -140°$ for the decay $\Lambda_b \to \Lambda \mu^+ \mu^-$. The blue plane represents the SM result and orange plane represents $Z'$ boson contribution.

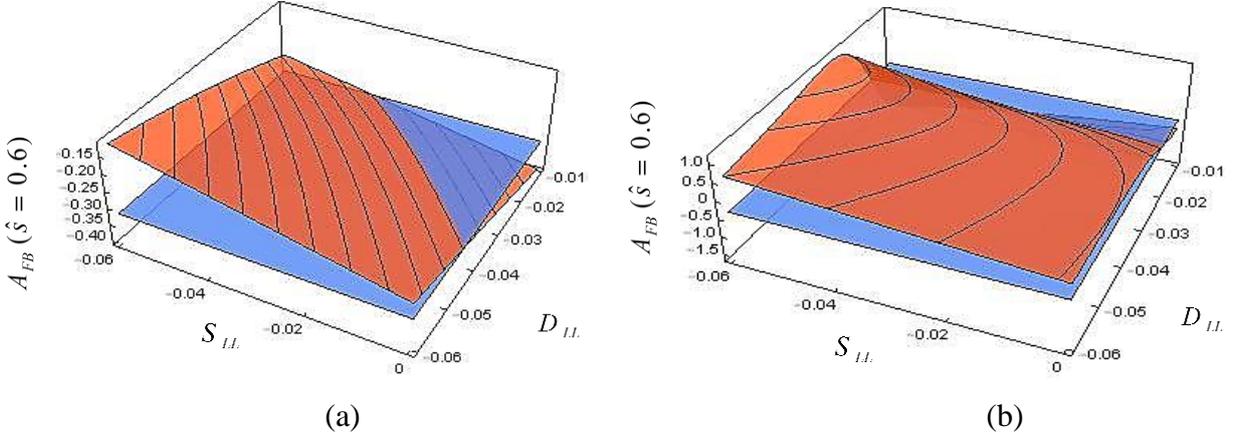

(a)          (b)

Fig.7: The dependence of $A_{FB}(\hat{s})$ on $S_{LL}$ and at $\hat{s} = 0.6$ with (a) $|B_{sb}| = 1.31 \times 10^{-3}$, $\phi_{sb} = -65°$ and (b) $|B_{sb}| = 5.7 \times 10^{-3}$, $\phi_{sb} = -140°$ for the decay $\Lambda_b \to \Lambda \tau^+ \tau^-$. The blue plane represents the SM result and orange plane represents $Z'$ boson contribution.

## 5. Conclusions

In this paper, we have studied the rare semileptonic $\Lambda_b \to \Lambda \ell^+ \ell^-$, $(\ell = e, \mu, \tau)$ decays in the SM as well as in a non-universal $Z'$ model. This non-universal $Z'$ model allows FCNC transitions at tree level, which gives a boost for the physical observables to compare with their SM values. Aliev *et al.* [78, 79] have discussed the decay width and lepton polarization for $\Lambda_b \to \Lambda \ell^+ \ell^-$ decays explicitly in the $Z'$ model. Their result shows an efficient tool for establishing new physics beyond the SM. Gutsche *et al.* [86] also presented a detailed study of observables for $\Lambda_b \to \Lambda \ell^+ \ell^-$ decays using the covariant quark model, and compared their results with others. In this paper, we have computed the variation of different physical observables with respect to $Z'$ couplings parameters within the kinematical region of $\hat{s}$. The effect of the $Z'$ mediated FCNCs enhance the differential decay rate, forward-backward asymmetry and lepton polarization asymmetries in these decay modes. Our results show deviations from the SM values, which is a signal of the presence of NP in these decays. The exploitation of the full data sets of the LHC experiments is a challenging task for both the theoretical and experimental communities. More precise measurements of $\Lambda_b \to \Lambda \ell^+ \ell^-$ decays will provide a powerful testing ground for the SM and possible NP models. We expect that the measurements of the observables will not only help us to find hints of NP but also provide a tool to determine the precise values of the parameters of the $Z'$ gauge boson.

**Acknowledgement**

We are grateful to the anonymous reviewers, whose suggestions have greatly improved the quality of our manuscript. D. Banerjee (IF140258) acknowledges the Department of Science and Technology (DST), Government of India for providing INSPIRE Fellowship. S. Sahoo acknowledges SERB, DST, Government of India for financial support (EMR/2015/000817).

**References**

1. T. Mannel and S. Recksiegel, *J. Phys. G* **24**, 979 (1998).
2. The CMS and LHCb Collaborations, *Nature* **522**, 68 (2015) [arXiv:1411.4413 [hep-ex]].
3. R. Aaij *et al.* (LHCb Collab.), *Phys. Rev. Lett.* **115,** 111803 (2015).
4. R. Aaij *et al.* (LHCb Collab.), *JHEP* **1207,** 133 (2012).
5. R. Aaij *et al.* (LHCb Collab.), *Phys. Rev. Lett.* **113,** 151601 (2014).
6. R. Aaij *et al.* (LHCb Collab.), *JHEP* **1406,** 133 (2014).
7. C. Patrignani *et al.* [Particle Data Group], *Chin. Phys. C* **40**, 100001 (2016).
8. R. Aaij *et al.* (LHCb Collab.), *JHEP* **1307,** 084 (2013).
9. W. S. Rou, R. S. Willey and A. Soni, *Phys. Rev. Lett.* **58**, 1608 (1987).

10. N. G. Deshpande and J. Trampetic, *Phys. Rev. Lett.* **60**, 2583 (1988); N. G. Deshpande, J. Trampetic and K. Ponose, *Phys. Rev. D* **39**, 1461 (1989).

11. C. S. Lim, T. Morozumi and A. I. Sanda, *Phys. Lett. B* **218**, 343 (1989)

12. B. Grinstein, M. J. Savage and M. B. Wise, *Nucl. Phys. B* **319**, 271 (1989).

13. C. Dominguez, N. Paver and Riazuddin, *Phys. Lett. B* **214**, 459 (1988).

14. T. M. Aliev, A. Özpinrci and M. Savci, *Phys. Rev. D.* **67**, 035007 (2003).

15. A. Ali, T. Mannel and T. Morozumi, *Phys. Lett. B* **273**, 505 (1991); A. Ali, G. F. Giudice and T. Mannel, *Z. Phys. C* **67**, 417 (1995).

16. C. Greub, A. Ioannissian and D. Wyler, *Phys. Lett. B* **346**, 149 (1995).

17. N. G. Deshpande, X.-G. He and J. Trampetic, *Phys. Lett. B* **367**, 362 (1996).

18. S. Bertolini, F. Borzumati, A. Masiero and G. Ridolfi, *Nucl. Phys. B* **353**, 591(1991).

19. F. Krüger and L. M. Sehgal, *Phys. Rev. D* **55,** 2799 (1997); *Phys. Rev. D* **56**, 5452 (1997).

20. S. Meinel and D. van Dyk, *Phys. Rev. D.* **94**, 013007 (2016).

21. C.-S. Huang and H.-J. Yan, *Phys. Rev. D* **59**, 114022 (1999); *Phys. Rev. D* **61**, 039901 (Erratum) (2000).

22. C.-H. Chen and C. Q. Geng, *Phys. Lett. B* **516**, 327 (2001) [hep-ph/0101201].

23. C.-H. Chen and C. Q. Geng, *Phys. Rev. D* **63**, 114024 (2001) [hep-ph/0101171]

24. C.-H. Chen and C. Q. Geng, *Phys. Rev. D* **64**, 074001 (2001) [hep-ph/0106193]

25. C.-H. Chen, C. Q. Geng and J. N. Ng, *Phys. Rev. D* **65**, 091502 (2000).

26. T. M. Aliev and M. Savci, *J. Phys. G* **26**, 997 (2000).

27. V. Berger, M. S. Berger, and R. J. N. Phillips, *Phys. Rev. D.* **52**, 1663 (1995).

28. J. Bernabeu, E. Nardi, D. Tommasini, *Nucl. Phys. B.* **409**, 69 (1993) [hep-ph/9306251].

29. P. Langacker, *Rev. Mod. Phys.* **81**, 1199 (2009) [arXiv:0801.1345 [hep-ph]].

30. J. Hewett and T. Rizzo, *Phys. Rep.* **183**, 193 (1989).

31. A. Leike, *Phys. Rep.* **317**, 143 (1999).

32. P. Langacker and M. Plümacher, *Phys. Rev. D* **62**, 013006 (2000).

33. S. Sahoo, *Indian J. Phys.* **80**, 191 (2008).

34. D. Feldman, Z. Liu and P. Nath, *Phys. Rev. Lett.* **97,** 021801 (2006) [hep-ph/0603039].

35. F. Abe *et al.* [CDF Collaboration], *Phys. Rev. Lett.* **77,** 438 (1996).

36. F. Abe *et al.* [CDF Collab.], *Phys. Rev. Lett.* **67**, 2418 (1991); *Phys. Rev. Lett.* **79**, 2192 (1997).

37. D. Abbaneo *et al.,* hep-ex/0212036.
38. V. Barger, C-W. Chiang, P. Langacker and H. S. Lee, *Phys. Lett. B* **580**, 186 (2004) [hep-ph/0310073].
39. S. Sahoo, C. K. Das, and L. Maharana, *Int. J. Mod. Phys. A* **26**, 3347 (2011).
40. S. Oda, N. Okda and D. Takahashi, *Phys. Rev. D* **92**, 015026 (2015).
41. M. Aaboud et al. (The ATLAS Collaboration), *Eur. Phys. J. C* **76**, 585 (2016).
42. V. Khachatryan *et al.* (CMS Collaboration), arXiv: 1701.01345 [hep-ex] (2017).
43. K. S. Babu, C. Kolda and J. March-Russell, *Phys. Rev. D* **54**, 4635 (1996) [hep-ph/9603212]; *Phys. Rev. D* **57**, 6788 (1998) [hep-ph/ 9710441].
44. S. Sahoo and L. Maharana, *Phys. Rev. D* **69**, 115012 (2004).
45. V. Barger, C-W. Chiang, P. Langacker and H. S. Lee, *Phys. Lett. B* **598**, 218 (2004).
46. V. Barger, C-W. Chiang, J. Jiang and P. Langacker, *Phys. Lett. B* **596**, 229 (2004).
47. A. Arhrib, K. Chung, C-W. Chiang and T-C. Yuan, *Phys. Rev. D* **73**, 075015 (2006) [hep-ph/0602175].
48. K. Cheung *et al.*, *Phys. Lett. B* **652**, 285 (2007) [ hep-ph/0604223].
49. Y. Nir and D. Silverman, *Phys. Rev. D* **42**, 1477 (1990).
50. G. Buchalla, A. J. Buras and M. E. Lautenbacher, *Rev. Mod. Phys.* **68**, 1125 (1996).
51. G. Bobeth, A. J. Buras F. Kruger, and J. Urban; *Nucl. Phys. B* **630**, 387 (2002).
52. A. Ali, arXiv:hep-ph/ 9606324.
53. A. Ali, P. Ball, L. T. Handoko, and G. Hiller, *Phys. Rev. D* **61**, 074024 (2000).
54. A. J. Buras and M. M*ü*nz, *Phys. Rev. D* **52**, 186 (1995).
55. A. S. Lim, T. Morozumi and A. I. Sanda, *Phys. Lett. B* **218**, 343 (1989)
56. P. J. O'Donnell, M. Sutherland and H. K. Tung, *Phys. Rev. D* **46**, 4091 (1992).
57. F. Kr ger and L. M. Sehgal, *Phys. Lett. B* **380**, 199 (1996).
58. A. K. Giri and R. Mohanta, *Eur. Phys. J. C* **45**, 151 (2006).
59. T. Mannel, W. Roberts and Z. Ryzak, Nucl. *Phys. B* **355**, 38 (1991).
60. C.-H. Chen, C. Q. Geng and J. N. Ng, arXiv:hep-ph/0210067.
61. Q. Chang and Y.-H. Gao, *Nucl. Phys. Proc. Suppl.* **115**, 263 (2003) [hep-ph/0210067].
62. Y. Li and J. Hua, *Eur. Phys. J. C* **71**, 1775 (2011) [arXiv:1107.0630 [hep-ph]].
63. H. Chen and H. Hatanaka, *Phys. Rev. D* **73**, 075003 (2006).
64. W. Chiang et al., *JHEP* **0608**, 075 (2006).
65. V. Barger et al., *Phys. Rev. D* **80**, 055008 (2009).

66. R. Mohanta and A. K. Giri, *Phys. Rev. D* **79**, 057902 (2009).

67. J. Hua, C. S. Kim, and Y. Li, *Eur. Phys. J. C* **69**, 139 (2010).

68. T. M. Aliev, K. Azizi, and M. Savci, *Phys. Rev. D* **81**, 056006 (2010).

69. Q. Chang, X. Q. Li, and Y. D. Yang, *JHEP* **1002**, 082 (2010) [arXiv:0907.4408 [hep-ph]].

70. M. Bona et al. [UTfit Collaboration], *PMC Phys. A* **3**, 6 (2009) [arXiv:0803.0659 [hep-ph]].

71. M. Bona *et al.*, [arXiv:0906.0953 [hep-ph]].

72. C.-H. Chen, *Phys. Lett. B* **683**, 160 (2010) [arXiv:0911.3479 [hep-ph]].

73. N. G. Deshpande, X.-G. He, and G. Valencia, *Phys. Rev. D* **82**, 056013 (2010) [arXiv:1006.1682 [hep-ph]].

74. J. E. Kim, M.-S. Seo, and S. Shin, *Phys. Rev. D* **83**, 036003 (2011) [arXiv:1010.5123 [hep-ph]].

75. P. J. Fox, J. Liu, D. Tucker-Smith, and N. Weiner, *Phys. Rev. D* **84**, 115006 (2011) [arXiv:1104.4127 [hep-ph]].

76. Q. Chang, R.-M. Wang, Y.-G. Xu, and X.-W. Cui, *Chinese Phys. Lett.* **28**, 081301 (2011).

77. X.-Q. Li, Y.-M. Li, and G.-R. Lin, *JHEP* **1205**, 049 (2012) [arXiv:1204.5250 [hep-ph]].

78. T. M. Aliev, K. Azizi, and M. Savci, *Phys. Lett. B* **718**, 566 (2012).

79. T. M. Aliev and M. Savci, *Nucl. Phys. B* **863**, 398 (2012).

80. M. Iwasaki *et al.* [Belle Collaboration], *Phys. Rev. D* **72**, 092005 (2005).

81. J. P. Lees *et al.* [BaBar Collaboration], [arXiv:1204.3993 [hep-ex]].

82. R. Aaij *et al.* [LHCb Collaboration], *Phys. Rev. Lett.* **108**, 181806 (2012).

83. R. Aaij *et al.* [LHCb Collaboration], *Phys. Rev. Lett.* **108**, 231801 (2012).

84. Q. Chang, X.-Q. Li and Y.-D. Yang, *J. Phys. G: Nucl. Part. Phys.* **41**, 105002 (2014).

85. M. A. Paracha, I. Ahmed and M. J. Aslam, *Prog. Theor. Exp. Phys.* **2015** 033B04 (2015).

86. T. Gutsche et al., *Phys. Rev. D* **87**, 074031 (2013).